\documentclass[10pt,aps,prd,amsmath,amssymb,nofootinbib,twoside,twocolumn,groupedaddress,superscriptaddress,
floats,floatfix,nolongbibliography]{revtex4-2}
\usepackage{etoolbox}
\apptocmd{\sloppy}{\hbadness 10000\relax}{}{}

\usepackage{kantlipsum}
\usepackage{comment}
\usepackage{graphicx,amssymb,amsmath,amsthm,amsfonts,epsfig,epsf,array,subfigure,appendix,booktabs}
\usepackage[usenames]{color}
\usepackage{xcolor}
\usepackage[unicode=true,pdfusetitle,
bookmarks=true,bookmarksnumbered=false,bookmarksopen=false, breaklinks=true,
colorlinks=true,citecolor=blue]
{hyperref}

\def\be{\begin{equation}}
\def\ee{\end{equation}}
\def\bea{\begin{eqnarray}}
\def\eea{\end{eqnarray}}
\def\bfla{\begin{flalign}}
\def\efla{\end{flalign}}
\def\nn{\nonumber}

\def\gsim{\, \rlap{$>$}{\lower 1.1ex\hbox{$\sim$}}\,}
\def\lsim{\, \rlap{$<$}{\lower 1.1ex\hbox{$\sim$}}\,}
\def\d{\mathrm{d}}
\def\x{\mathrm{x}}

\def\n{\mathrm{n}}
\def\s{\mathrm{s}}
\def\trho{\widetilde{\rho}}
\def\tpr{\widetilde{p}_r}
\def\tpt{\widetilde{p}_t}
\def\teta{\widetilde{\eta}}
\def\tdelta{\widetilde{\delta}}
\def\hrho{\widehat{\rho}}
\def\hpr{\widehat{p}_r}

\begin{document}

\title{\boldmath Self-similar collapse with elasticity}

\author{Jorge V.~Rocha}
\email{jorge.miguel.rocha@iscte-iul.pt}
\affiliation{Departamento de Matem\'atica, ISCTE--Instituto Universit\'ario de Lisboa, Avenida das For\c{c}as Armadas, 1649-026 Lisboa, Portugal}
\affiliation{Centro de Astrof\'isica e Gravita\c{c}\~ao--CENTRA, Instituto Superior T\'ecnico--IST, Universidade de Lisboa--UL, Av.\ Rovisco Pais 1, 1049-001 Lisboa, Portugal}
\affiliation{Instituto de Telecomunica\c{c}\~oes–-IUL, Avenida das For\c{c}as Armadas, 1649-026 Lisboa, Portugal}
\author{Diogo L. F. G.~Silva}
\email{diogo.l.silva@tecnico.ulisboa.pt}
\affiliation{Centro de Astrof\'isica e Gravita\c{c}\~ao--CENTRA, Instituto Superior T\'ecnico--IST, Universidade de Lisboa--UL, Av.\ Rovisco Pais 1, 1049-001 Lisboa, Portugal}

\date{\today}

\begin{abstract}
Critical collapse is a well-studied subject for a variety of self-gravitating matter. One of the most intensively examined models is that of perfect fluids, which have been used extensively to describe compact objects such as stars, as well as being of cosmological interest. However, neutron stars are believed to possess an elastic crust, thus departing from a perfect fluid body, and critical collapse with elastic materials is an entirely unexplored topic.
In this work, we employ a scale-invariant elastic matter model to study self-similar collapse with elasticity. As with perfect fluid models, we show that including elasticity allows for continuous self-similar configurations, which we determine numerically by solving the associated boundary value problem. The set of solutions is discrete and we focus on the fundamental mode, but also present some results for overtones.
Similarly to the perfect fluid case, the existence of a sonic point plays a central role. We find that the addition of elasticity, by either increasing the shear index $\s$ or decreasing the Poisson ratio $\nu$, leads to an increase in compressibility and can yield negative radial pressures around the sonic point. Simultaneously, the elastic longitudinal wave speed ceases to be constant, while the two possible transverse wave speeds grow further apart. The departure from the perfect fluid case can be so dramatic as to generate a second sonic point, which does not seem to be regular. This, in turn, imposes bounds on the elasticity parameters of the material.
This study represents the first step in the analysis of critical collapse with elastic materials.
\end{abstract}

\maketitle

\section{Introduction}

Critical phenomena occur in many branches of science, from condensed matter physics~\cite{Chaikin_Lubensky_1995} to complex systems~\cite{Dorogovtsev_2008}. In the context of gravitational physics, it arises at the verge of black hole formation from the collapse of self-gravitating bodies~\cite{Gundlach:2025yje}. This was first famously demonstrated by Choptuik~\cite{Choptuik:1992jv}, after suggestive theoretical work by Christodoulou~\cite{christodoulou:1986a, christodoulou:1986b, Christodoulou:1991yfa} analyzing the dynamics of a massless scalar field coupled to Einstein gravity. 

Choptuik~\cite{Choptuik:1992jv} studied a matter system composed of a minimally coupled scalar field in spherical symmetry, finding three important characteristics: power-law scaling, universality, and self-similarity. 
Specifically, by numerically evolving different one-parameter families of initial data, Choptuik verified the existence of a threshold value for the parameter separating the two possible end-states (total dispersion or black hole formation). Furthermore, when tuning the parameter slightly above the threshold value, the mass of the black hole formed follows a power law whose critical exponent is independent of the initial data family chosen. Finally, precisely at the threshold value the system evolves in a (discretely) self-similar manner. These results have been extensively verified over the years~\cite{Gundlach:1993tn,Garfinkle:1994jb,Hamade:1995ce,Gundlach:1995kd,Gundlach:1996eg,Brady:1997fj}.
Similar phenomena have been observed with perfect fluids~\cite{evans:1994,Koike:1995jm,Maison:1995cc,Neilsen:1998qc,Koike:1999eg}, Yang-Mills fields~\cite{Choptuik:1996yg,Gundlach:1996je}, massless scalar electrodynamics~\cite{Gundlach:1996vv}, and axisymmetric pure gravitational waves~\cite{Abrahams:1993wa}, to name a few. 

Critical collapse of perfect fluids~\cite{evans:1994,Koike:1995jm,Maison:1995cc,Neilsen:1998qc,Koike:1999eg} stands out in, at least, two respects: (i) such matter is appropriate to describe most stars, as well as many other celestial objects and astrophysical structures, and (ii) arguably they provide the simplest $3+1$-dimensional matter model that displays self-similarity of the continuous kind, in contrast with the minimally coupled massless scalar model, which enjoys discrete self-similarity.

Concerning the second point, one can take advantage of continuous self-similarity (CSS) in combination with spherical symmetry to cast the equations of motion as a system of ordinary differential equations (ODEs), as first pointed out in~\cite{Cahill:1970ew} and also explored in~\cite{Ori:1989ps}. The problem is then amenable to a dynamical systems' treatment, as advocated by Koike, Hara and Adachi~\cite{Koike:1999eg},
the self-similar solution being identified with a fixed point of the system, while the Lyapunov exponents of linear perturbations are simply related to the critical
exponent, borrowing renormalization group ideas~\cite{Koike:1995jm}. 

Regarding point (i) above, the material most stars are made of is well approximated by a perfect fluid. However, this does not seem to be the case for neutron stars, for which observations indicate the need to consider stellar structures that comprise an elastic solid crust~\cite{Chamel:2008ca}. One such example is that of pulsar glitches, sudden deviations in the body's angular frequency.\footnote{Elasticity is also relevant in potential explanations of giant flares from soft gamma repeaters~\cite{Chamel:2008ca}.}
The effect of crustal elasticity on gravitational waves generated by neutron star binary mergers is not too encouraging for detectability~\cite{Penner:2011br}, and its impact on tidal deformability is small~\cite{Penner:2011pd,Gittins:2020mll}.

Universality properties are extremely useful if one aims to make general predictions. Previous work has employed this characteristic of critical collapse in a cosmological context to derive the initial mass distribution of primordial black holes~\cite{Niemeyer:1997mt,Green:1999xm,Hawke:2002rf,Musco:2004ak,Musco:2008hv} and its effect on gravitational wave detection~\cite{Franciolini:2022tfm,Escriva:2022bwe}.
Nevertheless, a word of caution is in order. It is by now well-known that universality does not extend across different matter models~\cite{Maison:1995cc}, i.e., it applies only to different families of initial data within the same matter model, at most.  Moreover, there are indications that universality might also be lost once the assumption of spherical symmetry is dropped~\cite{Choptuik:2003ac, Baumgarte:2023tdh,Marouda:2024epb}.

When elasticity is added to a self-gravitating body its properties are expected to change. This has been demonstrated in stationary neutron star configurations ---for instance, \cite{Pereira:2020cmv} considered hybrid star models with an elastic quark innermost phase, in addition to the usual crust, finding that this causes relative radius deviations at the percent level--- but so far it has not been studied in the context of critical collapse.
Could the introduction of elasticity destroy the critical behavior observed in perfect fluids? If criticality is preserved, does it still remain of CSS type? How does it modify the critical exponent? These are the kind of questions we aim to answer.

To pursue this avenue, we must consider elastic generalizations of perfect fluids in the context of general relativity. 
The first complete self-consistent formulation of relativistic elasticity was developed by Carter and Quintana~\cite{Carter:1972} with the introduction of convective frames, i.e, frames directed along the 4-velocity. 
Karlovini and Samuelsson~\cite{Karlovini:2003} used exclusively tensors defined on either physical or material spaces and their push-forwards and pull-backs, thereby removing the need for convective frames. 
Beig and Schmidt~\cite{Beig:2003} developed a theory of elasticity starting from a Lagrangian covariant under spacetime diffeomorphisms. 
Brown~\cite{Brown:2020pav} presented Lagrangian formulation of relativistic elasticity.
More recently, Alho \textit{et al.} \cite{Alho:2023ris} (see also~\cite{Alho:2021sli}) explored a new Eulerian approach for relativistic elastic bodies in spherical symmetry, which we will adopt.

In this paper we initiate a program to investigate critical collapse with relativistic elastic matter.
We follow the Koike-Hara-Adachi approach~\cite{Koike:1999eg}. We will show that elastic generalizations of the perfect fluid polytropes considered therein also allow continuous self-similar collapsing solutions.
The inclusion of elasticity enlarges the dynamical variables to solve for, so in this work we restrict ourselves to obtaining (families of) continuous self-similar solutions and studying their properties as the elasticity parameters are varied. We emphasize that here we do not study perturbations of these backgrounds, which allows one to determine whether the solutions obtained are critical and, if so, to compute their critical exponent. That is the subject of a forthcoming paper.

\section{Continuous self-similar collapse}

\subsection{Metric ansatz}

We consider a spherically symmetric spacetime, for which the metric can be generically written as
\be
    \d s^2= -\alpha^2(t,r)\, \d t^2+ \beta^2(t,r)\, \d r^2+ r^2\, \d \Omega_2^2\,,
\ee
with $\alpha(t,r)$ and $\beta(t,r)$ yet undetermined metric functions, and $\mathrm{d}\Omega_2^2 = d\theta^2 + \sin^2 \theta\, d\phi^2$ the line element of the unit 2-sphere. 
We impose continuous self-similarity by demanding the existence of a homothetic vector field $\xi$, under whose flow the metric preserves its form up to a scale transformation, 
\be
    \label{eqn:SSrelation}
    \mathcal{L}_{\xi}g_{ab}= 2\, g_{ab}\,,
\ee
Choosing $\xi=t\,\partial_t+ r\, \partial_r$, and changing to new coordinates adapted to self-similarity, $(\tau,x)$, defined through
\be
\label{eq:taux}
    \tau=-\ln{\left( -\frac{t}{l} \right)}\,, \quad 
    x= \ln{\left( -\frac{r}{t} \right)}\,,
\ee
with $l$ being an arbitrary scale factor, Eq.~(\ref{eqn:SSrelation}) is verified if and only if $\alpha$ and $\beta$ are functions of $x$ only, i.e., if $\alpha\equiv \alpha(x)$ and $\beta\equiv \beta(x)$.

It is convenient to exchange the metric functions $\alpha$ and $\beta$ for new ones, $A$ and $N$, defined as
\be
    A = \beta^2, \quad N = \frac{\alpha}{\beta\, e^x}\,.
\ee
With this choice, the metric takes the form
\be
    \label{eqn:SSmetric}
    \mathrm{d}s^2= l^2\, \mathrm{e}^{2\, x -2\, \tau}\, \mathrm{d}\hat{s}^2\,,
\ee
with the conformal metric $\mathrm{d}\hat{s}^2$ given by
\be
    \d \hat{s}^2=  -(N^2- 1)\, A\, \d \tau^2- 2\, A\, \d \tau\, \d x + A\, \d x^2 + \d \Omega_2^2\,.
\ee

\subsection{Relativistic elasticity formalism}

Let us now turn to the matter sector.
We assume a material represented by an elastic matter model with a spherically symmetric stress-energy tensor
\be
    T=\left[ \left(\rho+ p_r \right)\, u_a\, u_b+ p_r\, g_{ab} \right]\, \d \x^a\, \d \x^b + p_t\, r^2\, \mathrm{d}\Omega_2^2\,,
\ee
where $\x^a=(t,r)$.
The 4-velocity $u$ of the matter particles is chosen consistently with spherical symmetry, having only the following two non-vanishing components:
\be
\label{eq:utur}
    u^t=\frac{1}{\alpha\, \sqrt{1- V^2}}\,,
    \quad 
    u^r=\frac{V}{\beta\, \sqrt{1-V^2}}\,,
\ee
with $V$ being the 3-velocity of fluid particles.

Similarly to what happens with perfect fluids, one can show that continuous self-similarity, as defined by Eq.~\eqref{eqn:SSrelation}, imposes
\begin{align}
\begin{split}
\label{eq:CSSmatter}
	\mathcal{L}_{\xi}u^a &= - u^a\,, \qquad\;\,
	\mathcal{L}_{\xi}\rho \:= - 2\rho\,, \\
	\mathcal{L}_{\xi}p_r &= - 2p_r\,, \qquad
	\mathcal{L}_{\xi}p_t = - 2p_t\,.
\end{split}
\end{align}
The first of these conditions implies that, in the self-similar coordinates $(\tau,x)$, the components of the 4-velocity can be generic functions of $x$ times $e^\tau$. In fact, from Eqs.~\eqref{eq:utur} one has
\bea
    \label{eqn:SS4velpar1}
    u^{\tau} &=& \frac{e^{\tau-x}}{l}\, \left( \frac{1}{N\, \sqrt{A}\, \sqrt{1-V^2}} \right)\,, \\ 
    \label{eqn:SS4velpar2}
    u^x &=& \frac{e^{\tau-x}}{l}\, \left( \frac{1+N\, V}{N\, \sqrt{A}\, \sqrt{1-V^2}} \right)\,,
\eea
so we conclude that $V$ can be a function only of $x$.

For the elastic reference state, we consider a {\em flat} rest configuration associated with the 3-submanifold describing the elastic matter, with a material metric of the form
\be
    \gamma_{IJ}\,\d \mathrm{X}^I\,\d \mathrm{X}^J =  \d R^2+ R^2\,\left( \d \Theta^2+ \sin^2{\Theta}\,\d \varPhi^2\right)\,,
\ee
with $\mathrm{X}^I=(R,\Theta,\varPhi)$ denoting the coordinates of the 3-submanifold. The requirement of a diffeomorphism between the physical spacetime and the reference state induces the choice $R=R(\x)$, $\Theta=\theta$ and $\varPhi=\phi$. 

The equation of state is established in terms of the eigenvalues, $n_r$ and $n_t$, of the deformation operator, defined by the projection of the material's 3-metric on the spacetime metric pushed-forward onto the reference frame ~\cite{Alho:2023ris}. 
More precisely, the normalized particle number density $\delta=n_r\, n_t^2$, and the normalized average number of particles $\eta=n_t^3$, are given, in the Schwarzschild coordinates we have adopted, by
\be
    \!\!\!
    \delta(t,r)=\frac{\sqrt{1-V^2}}{\beta}\, \left( \frac{R}{r} \right)^2\, \partial_rR,
    \quad 
    \eta(t,r)=\left( \frac{R}{r} \right)^3.
\ee


It is more convenient to express the energy density and pressures in terms of $\delta$ and $\eta$ ---which is made explicit henceforth by a hat---, rather than directly as functions of the spacetime coordinates $(t,r)$. 
The variables $\delta$ and $\eta$ are more useful to us than the material's energy density and pressures.
An equation of state is then a function $\hrho\equiv \hrho(\delta,\eta)$, and the radial and tangential pressures follow from~\cite{Alho:2023ris}
\begin{align}
    &\hpr= \delta\, \frac{\partial \hrho}{\partial \delta}- \hrho\,, \label{eq:pr}\\
    &\widehat{p}_t= \hpr+ \frac32\, \eta\, \frac{\partial \hrho}{\partial \eta}\,. \label{eq:pt}
\end{align}

Recall also that $p = k \rho$ is the only barotropic equation of state compatible with self-similarity~\cite{Ori:1989ps}. However, the inclusion of elasticity allows for other possibilities, as we now briefly discuss.

\bigskip
\subsection{Scale-invariant matter model}

Recently it was shown that scale-invariant elastic matter models in spherical symmetry generically verify~\cite{Alho:2023mfc}
\be
    \hrho= \frac{\n}{3} \left( \hpr+ 2\, \widehat{p}_t \right)\,,
\ee
where $n>0$ is the polytropic index. Together with Eqs.~(\ref{eq:pr}-\ref{eq:pt}), this determines the form of the material's energy density to be
\be
	\hrho = \frac{\n^2 k}{\n+1} \, \eta^{1+\frac{1}{\n}} h\left(\frac{\delta}{\eta}\right)\,,
\ee
with $h$ a free function and $k$ a positive constant. 
By taking a power-law expression for $h$, consistent with the isotropic state and linear elasticity conditions, the following form was obtained for the energy density~\cite{Alho:2023mfc}:
\begin{widetext}
\be
\label{eq:ScaleInvariantRho}
    \hrho= \frac{\n^2\, k}{\n+1} \, \eta^{1+ \frac{1}{\n}}\, \left[ 1 - \frac{\n+1}{\n} \left( 1- 3\, \frac{\s}{\n}\, \frac{1-\nu}{1+\nu} \right) \left( 1 - \frac{\delta}{\eta} \right)  -  3\, \frac{\s^2}{\n^2}\,\frac{\n+1}{\s+1}\,  \frac{1- \nu}{1+ \nu}  \, \left( 1- \left (\frac{\delta}{\eta} \right)^{1+ \frac{1}{\s}} \right) \right],
\ee
\end{widetext}
with $\s$ the shear index and $\nu$ the Poisson ratio.
The associated pressures can be obtained from Eqs.~(\ref{eq:pr}-\ref{eq:pt}).

The perfect fluid case with a linear equation of state is recovered for $\n=\s$ and $\nu=1/2$, which yields $\hrho(\delta)=\n\, \hpr(\delta)=\n\, \widehat{p}_t(\delta)=\frac{\n^2 k}{\n+1}\, \delta^{1+\frac{1}{\n}}$, a $\eta$-independent expression that also allows one to identify $\gamma=1+\frac{1}{\n}$ as the adiabatic index.
In this case, the choice $\n=3$ returns the radiation fluid, for which the critical solution was first obtained by Evans and Coleman~\cite{evans:1994}.

By introducing the rescaled functions
\begin{eqnarray}
    \tdelta^{1+ \frac{1}{\n}}&= 4\,\pi\,k\,r^2\,\beta^2\,\delta^{1+ \frac{1}{\n}}\,,\label{eq:deltatilde}\\
    \teta^{1+ \frac{1}{\n}}&= 4\,\pi\,k\,r^2\,\beta^2\,\eta^{1+ \frac{1}{\n}}\,,\label{eq:etatilde}
\end{eqnarray}
the self-similarity constraints on the energy density and pressures, Eqs.~\eqref{eq:CSSmatter}, are satisfied if and only if $\tdelta$ and $\teta$ are functions only of $x$. Likewise we introduce 
\be
\label{eq:tilded_hatted}
    \trho=4\, \pi\, k\, r^2\, \beta^2\, \hrho\,, 
    \qquad
    \widetilde{p}_i=4\, \pi\, k\, r^2\, \beta^2\, \widehat{p}_i\, \;\; (i=r,t)\,.
\ee
Using the tilded set of variables, the wave speeds are defined as~\cite{Alho:2023ris}
\begin{align}
    c_L^2 &= \frac{\tdelta\, \partial_{\tdelta}\tpr}{\trho+\tpr}\,,
    \label{eq:cL}\\
    c_T^2 &= \frac{\tpt-\tpr}{\left( \trho+\tpt \right)\, \left( 1- \tdelta^2/\teta^2 \right)}\,,
    \label{eq:cT}\\
    \widetilde{c}_T^2 &= \frac{\tpr-\tpt}{\left( \trho+\tpr \right)\, \left( 1- \teta^2/\tdelta^2 \right)}\,,
    \label{eq:cTtilde}
\end{align}
where $c_L$ is the speed of the longitudinal wave propagating radially, $c_T$ is the speed of the transverse wave propagating radially, and $\widetilde{c}_T$ is the speed of a transverse wave that propagates tangentially and oscillates along the radial direction.

\subsection{Equations of motion}

We have seen that the choice of a self-similar spacetime and matter model requires that the five functions $A, N, V, \tdelta$ and $\teta$ depend only on $x$. In this case, the field equations become ordinary differential equations. Denoting derivatives with respect to $x$ with a prime, the Einstein field equations directly yield
\begin{align}
    \label{eqn:EOMA1}
    &\frac{A'}{A}= 1- A+ \frac{2}{1-V^2}\, \left( \trho+ V^2\, \tpr \right)\,,\\
    \label{eqn:EOMA2}
    &\frac{A'}{A}= -\frac{2\, N\, V}{1-V^2}\,\left( \trho+ \tpr \right)\,,\\
    \label{eqn:EOMN}
    &\frac{N'}{N}= -2+ A- (\trho- \tpr)\,.
\end{align}
Additionally, conservation of energy and momentum give
\begin{widetext}
\begin{flalign}
    &(1+N\,V)\, (\trho+ \tpr)\, \frac{\tdelta'}{\tdelta} - \frac23\, (1+ N\,V)\, (\tpr- \widetilde{p}_t)\, \frac{\teta'}{\teta} 
     + \frac{N+V}{1-V^2}\, (\trho+ \tpr)\,V' =
    \nn \\
    & \qquad\qquad\qquad\qquad\qquad\qquad\qquad 
    = \frac32\, N\, V\, \left( \frac{\n-2}{\n}\, \trho+ \tpr \right) - \frac12\, A\, N\, V\, ( 3\, \trho+ \tpr) + N\, V\, \trho\, (\trho- \tpr)\,,
    \label{eqn:EOMBianchi1}
    \\
    &(N+V)\, c_L^2\, \left( \trho+ \tpr \right)\, \frac{\tdelta'}{\tdelta}- (N+ V)\, \left( c_L^2\, \left( \trho+ \tpr \right)- \frac{\n+ 1}{\n}\, \tpr \right)\, \frac{\teta'}{\teta} + \frac{1+ N\, V}{1-V^2}\, (\trho+ \tpr)\, V' =
    \nn \\
    & \qquad\qquad\qquad\qquad\qquad\qquad\qquad 
    = N\, (\trho- \tpr)\, \tpr + \frac{1}{2\, \n}\, N\, \left[ (\n+6)\, \trho+ \n\, \tpr \right]- \frac12\, A\, N\, (\trho+ 3\, \tpr)\,,
    \label{eqn:EOMBianchi2}
\end{flalign}
\end{widetext}
while the elastic relations connecting the matter variables $\tdelta$ and $\teta$ (see~\cite{Alho:2023ris}) 
yield
\begin{flalign}
    \label{eqn:EOMeta1}
    &\!\!\!\!\frac{\teta'}{\teta}= \left( \frac{3\, \sqrt{A}}{\sqrt{1-V^2}} \right)\, \frac{\tdelta}{\teta} -\frac{\n+3}{\n+1}- \frac{2\, \n\, N\, V\, (\trho+ \tpr)}{\left(\n+1\right)\, \left(1-V^2\right)}\,,\\
    \label{eqn:EOMeta2}
    &\!\!\!\!\frac{\teta'}{\teta}= -\left( \frac{3\, \sqrt{A}\,N\, V}{\sqrt{1-V^2}} \right)\, \frac{\tdelta}{\teta}- \frac{2\, \n\, N\, V\, (\trho+ \tpr)}{\left( \n+1 \right)\, \left(1-V^2\right)}\,.
\end{flalign}

\bigskip
Among the preceding seven relations, only five are independent. Both pairs of Eqs.~\eqref{eqn:EOMA1} and~\eqref{eqn:EOMA2} and Eqs.~\eqref{eqn:EOMeta1} and~\eqref{eqn:EOMeta2} can be used to obtain algebraic relations,
\begin{flalign}
    \label{eqn:algebraic1}
    & 1- A+ \frac{2\,\left[ (1+NV)\trho+V(N+V)\, \tpr \right]}{1-V^2} = 0\,,\\
    \label{eqn:algebraic2}
    & \frac{3\, \sqrt{A}(1+NV)}{\sqrt{1-V^2}}\, \frac{\tdelta}{\teta} = \frac{\n+3}{\n+1}\,.
\end{flalign}
One might use these constraints to eliminate two variables, but we will not do so. As in~\cite{Koike:1995jm}, we will solve the full set of ODEs for our five variables and use Eqs.~\eqref{eqn:algebraic1} and \eqref{eqn:algebraic2} to check the numerical precision of our results.

Substituting Eq.~\eqref{eqn:EOMeta2} in Eqs.~\eqref{eqn:EOMBianchi1} and \eqref{eqn:EOMBianchi2} reduces both to a set of two coupled ordinary differential equations for $\tdelta$ and $V$,
\begin{widetext}
\begin{flalign}
    \label{eqn:EOMBianchif1}
    &(1+ N\,V)\, (\trho+ \tpr)\, \frac{\tdelta'}{\tdelta}+ \frac{N+V}{1-V^2}\, (\trho+ \tpr)\,V' =
    \nn \\
    & \qquad\qquad
    = \frac32\, N\, V\, \left( \frac{\n-2}{\n}\, \trho+ \tpr \right)- \frac{A\, N\, V}{2}\, (3\, \trho+ \tpr) + N\, V\, \trho\, (\trho- \tpr)- \frac23 (1+ N\, V)\, (\tpr- \widetilde{p}_t)\, \Lambda\,,
    \\
    \label{eqn:EOMBianchif2}
    &(N+V)\, c_L^2\, \left( \trho+ \tpr \right)\, \frac{\tdelta'}{\tdelta}+ \frac{1+ N\, V}{1-V^2}\, (\trho+ \tpr)\, V'=
    \nn \\
    & \qquad\qquad
    = N\, (\trho- \tpr)\, \tpr+ \frac{N}{2\, \n}\, \left[ (\n+6)\, \trho+ \n\, \tpr \right] -\frac{A\, N}{2}\, (\trho+ 3\, \tpr)- (N+ V)\, \left( c_L^2\, \left( \trho+ \tpr \right)- \frac{\n+1}{\n}\, \tpr \right)\, \Lambda\,,
\end{flalign}
\end{widetext}
where
\be
    \Lambda= \left( \frac{3\, \sqrt{A}\, N\, V}{\sqrt{1-V^2}} \right)\, \frac{\tdelta}{\teta}+ \frac{2\, \n\, N\, V\, \left( \trho+ \tpr \right)}{\left( \n+1 \right)\, \left(1-V^2\right)}\,.
\ee

Note that the relations~\eqref{eq:tilded_hatted}, (\ref{eq:ScaleInvariantRho}-\ref{eq:etatilde}) and~\eqref{eq:pr}, allow one to express $\trho$ and $\tpr$ in terms of $\tdelta$, $\teta$ and $A$. 
Thus, Eqs.~\eqref{eqn:EOMA2}, \eqref{eqn:EOMN}, \eqref{eqn:EOMeta2}, \eqref{eqn:EOMBianchif1} and \eqref{eqn:EOMBianchif2} constitute a system of five first-order ordinary differential equations for five functions: $A(x)$, $N(x)$, $\teta(x)$, $V(x)$, $\tdelta(x)$. The problem involves three free parameters: $\n$, $\s$, $\nu$. Note that $k$ only determines an overall scale and was effectively absorbed by the transformations~\eqref{eq:deltatilde} and~\eqref{eq:etatilde}.

\section{Self-similar solutions for 
collapse with elastic matter}
\vspace{-0.2cm}

Solutions of the dynamical system determined by Eqs.~\eqref{eqn:EOMA2}, \eqref{eqn:EOMN}, \eqref{eqn:EOMeta2}, \eqref{eqn:EOMBianchif1} and \eqref{eqn:EOMBianchif2} yield self-similar spacetimes supported by the scale-invariant elastic matter model considered.

\vspace{-0.3cm}
\subsection{The sonic point}
\vspace{-0.2cm}

Contrary to Eqs~\eqref{eqn:EOMA2}, \eqref{eqn:EOMN} and \eqref{eqn:EOMeta2}, the system of equations~\eqref{eqn:EOMBianchif1} and \eqref{eqn:EOMBianchif2} is still coupled. The latter can be written in the form
\be
    \label{eqn:couplematrix}
    \begin{bmatrix}
        \mathrm{a} & \mathrm{b}\\
        \mathrm{c} & \mathrm{d}
    \end{bmatrix}
    \begin{bmatrix}
        \tdelta'\\
        V'
    \end{bmatrix}
    =
    \begin{bmatrix}
        \mathrm{e}\\
        \mathrm{f}
    \end{bmatrix}\,,
\ee
with
\begin{flalign}
    &\mathrm{a}=(1+ N\,V)\, \frac{\trho+\tpr}{\tdelta}\,,
    \quad\quad
    \mathrm{b}=\frac{N+V}{1-V^2}\, (\trho+ \tpr)\,,
    \label{eq:ab}\\
    &\mathrm{c}=(N+ V)\, c_L^2\, \frac{\trho+\tpr}{\tdelta} \,,
    \quad\;\,
    \mathrm{d}=\frac{1+ N\, V}{1-V^2}\, (\trho+ \tpr)\,,
    \label{eq:cd}\\
        &\mathrm{e}=\frac32\, N\, V\, \left( \frac{\n-2}{\n}\, \trho+ \tpr \right)- \frac{A\, N\, V}{2}\, ( 3\, \trho+ \tpr )\\
        &+ N\, V\, \trho\, (\trho- \tpr)- \frac23 (1+ N\, V)\, (\tpr- \widetilde{p}_t)\, \Lambda\,,\nn\\
        &\mathrm{f}=N\, (\trho- \tpr)\, \tpr+ \frac{N}{2\, \n}\, \left[ (\n+6)\, \trho+ \n\, \tpr \right]\\
        &- \frac{A N}{2} (\trho+ 3\, \tpr) - (N+ V) \left( c_L^2 \left( \trho+\tpr \right) - \frac{\n+1}{\n}\, \tpr \right)\, \Lambda\,.\nn
\end{flalign}
Whenever $\mathrm{a}\,\mathrm{d}-\mathrm{b}\,\mathrm{c}\neq 0$ the system of Eqs.~\eqref{eqn:EOMBianchif1} and~\eqref{eqn:EOMBianchif2} can be explicitly decoupled,
\be
    \label{eqn:EOMcoupled}
    \tdelta'= \frac{\mathrm{d}\, \mathrm{e} - \mathrm{b}\, \mathrm{f}}{\mathrm{a}\, \mathrm{d} - \mathrm{b}\, \mathrm{c}}\,, \qquad
    V'=\frac{\mathrm{a}\, \mathrm{f} - \mathrm{c}\, \mathrm{e}}{\mathrm{a}\, \mathrm{d} - \mathrm{b}\, \mathrm{c}}\,.
\ee

Equations~\eqref{eqn:EOMcoupled}, together with~\eqref{eqn:EOMA2}, \eqref{eqn:EOMN} and \eqref{eqn:EOMeta2}, form a system of ordinary differential equations that can be straightforwardly integrated numerically, except at singular points of Eq.~\eqref{eqn:couplematrix}, where
\be
    \mathrm{det}
    \begin{bmatrix}
        \mathrm{a} & \mathrm{b}\\
        \mathrm{c} & \mathrm{d}
    \end{bmatrix}=0\,.
\ee
These events are called sonic points. They occur in self-similar collapse of perfect fluids, and adding elasticity to the model preserves them. There exists a residual gauge freedom that allows one to shift the self-similar coordinate, and we use this to place the singular point at $x=0$.

Sonic points correspond to points where an observer with fixed self-similar coordinate $x$ sees the particles of the collapsing star moving with velocity equal to the material's wave speed~\cite{Harada:2001hk}.
The former is obtained from the projection of the particles' 4-velocity, Eqs.~(\ref{eqn:SS4velpar1}-\ref{eqn:SS4velpar2}), onto the 4-velocity of the observer,
\be
    -\left(u_{obs}\right)_{\alpha}\, \left( u_{par} \right)^{\alpha}= \frac{1}{\sqrt{1-v^2}}\,,
\ee
where $v$ is the 3-velocity of the particles as seen by an observer at fixed $x$, and $u_{obs}$ is the 4-velocity of the fixed observer,
\be
    \label{eqn:SS4velobs}
    u_{obs}^{\tau}=\frac{e^{\tau-x}}{l \sqrt{A\,(N^2-1)}}\,.
\ee
It follows that
\be
    v=\sqrt{1- \frac{\left(N^2-1\right)\, \left(1-V^2\right)}{\left( N+V \right)^2}}\,.
    \label{eq:stellarvelocity}
\ee
It should be noted that $u_{obs}^{\tau}$ is not well defined for $N^2<1$, owing to the exchange of character among the $\tau$ and $x$ coordinates in such regions of spacetime, as can be seen from the self-similar metric of Eq.~(\ref{eqn:SSmetric}). Nevertheless, the velocity $v$ can be analytically continued across $N=1$, becoming in that case larger than the speed of light.

While the system of ODEs breaks down at the sonic point, we demand smoothness of the solution at all points. Thus, the solution is obtained, at the sonic point, using a local series expansion. The aforementioned requirements on the solution also entail that four relations be met at the sonic point. Two correspond to the algebraic relations verified at all points, respectively Eqs.~\eqref{eqn:algebraic1} and \eqref{eqn:algebraic2}. Two others are established by
\begin{eqnarray}
    \label{eqn:algebraicSP1}
    \mathrm{a}\, \mathrm{d} - \mathrm{b}\, \mathrm{c}=0\,,\\
    \label{eqn:algebraicSP2}
    \mathrm{a}\, \mathrm{f} - \mathrm{e}\, \mathrm{c}=0\,,
\end{eqnarray}
imposing analyticity of the system at the sonic point~\cite{Ori:1989ps,Koike:1995jm,Harada:2001hk}. This set of conditions reduces the five dimensional space of solutions, at the sonic point, to a one-parameter family. The free parameter is taken to be $V_0$, the particle 3-velocity, in Schwarzschild coordinates, at the sonic point.

\subsection{$x\to-\infty$ asymptotic behavior}

The center of the collapsing body, to the past of the accumulation point $t=r=0$, corresponds to $x\to-\infty$, as can be seen from~\eqref{eq:taux}. Aiming for CSS solutions that are regular everywhere except at the accumulation point, one must impose regularity also as $x\to-\infty$. When this limit is taken, the function $N$ diverges.
As such, we follow~\cite{Hara:1996mc} and introduce a new dependent variable, $M$, defined as
\be
    M=N\,V\,.
\ee
Making this replacement on the set of ODEs~\eqref{eqn:EOMA2}, \eqref{eqn:EOMN}, \eqref{eqn:EOMeta2}, \eqref{eqn:EOMBianchif1} and \eqref{eqn:EOMBianchif2}, one finds the fixed point
\begin{equation}
\begin{split}
    &A^*=1\,, \quad M^*= -\frac{2\,\mathrm{n}}{3\, (1+\mathrm{n})}\,, \\
    &\teta^*=0\,, \quad \tdelta^*=0\,, \quad V^*=0\,,
\end{split}
\end{equation}
where, additionally,
\begin{equation}
    \lim_{x\to -\infty}\frac{\tdelta}{\teta}=1\,.
\end{equation}

Application of perturbative analysis around this fixed point yields the asymptotic behavior
\begin{equation}
\begin{split}
    &A(x)\sim 1+ A_{-\infty}\, e^{2\,x}, 
    \quad 
    N(x)\sim N_{-\infty}\, e^{-x},
    \\
    &\teta(x)\sim \teta_{-\infty}\, e^{\frac{2\, \n}{\n+1} x},
    \qquad\;
    \tdelta(x)\sim \tdelta_{-\infty}\, e^{\frac{2\, \n}{\n+1} x}, 
    \\
    &V(x)\sim V_{-\infty}\, e^{x},
\end{split}
\end{equation}
where the constants $A_{-\infty}$, $N_{-\infty}$, $\teta_{-\infty}$, $\tdelta_{\infty}$ and $V_{-\infty}$ are constrained by
\begin{flalign}
    &A_{-\infty}=\frac{2\,\n^2}{3\, \left(\n+1 \right)} \, \tdelta_{-\infty}^{1+\frac{1}{\n}}\,, \\ 
    &N_{-\infty}\, V_{-\infty}= -\frac{2\, \n}{3\, (\n+1)}\,,\\
    &\teta_{-\infty}=\tdelta_{-\infty}\,.
\end{flalign}
From these relations, one sees that only two remain free, which we choose to be $\tdelta_{-\infty}$ and $V_{-\infty}$.

We remark that this fixed point exactly matches the one found in the perfect fluid case. This is to be expected, as elastic matter models are isotropic in the center and, thus, locally similar to perfect fluids.

\subsection{Numerical approach to 
solve the boundary value problem}
\label{sec:NumApproach}

Under the assumption that there is a unique sonic point, no other singular points of the system of ODEs exist, and so the requirements on the parameters at the sonic point and at the center casts the system as a boundary value problem in the domain ${\cal D}=\{x\in\mathbb{R}:-\infty<x\leq0\}$. 
Demanding the solution to be continuous and differentiable everywhere in ${\cal D}$ reduces the number of solutions to a discrete set. One may discriminate solutions by their corresponding value of $V_0$, for example.
In the following, we further restrict the solutions by requiring that they have only one sonic point. Solutions with multiple sonic points are most likely singular, since each sonic point introduces further regularity constraints, thus reducing even more the number of free parameters~\cite{Koike:1999eg}.

We employ a shooting method to obtain regular solutions of the boundary value problem. Recall that we have three free parameters ($V_0,\tdelta_{-\infty},V_{-\infty}$), one of them at the rightmost boundary, the other two at the leftmost end. Demanding continuity of all functions at an (arbitrary) intermediate point imposes three conditions, taking into account the algebraic constraints~(\ref{eqn:algebraic1}-\ref{eqn:algebraic2}). The ODE system being of first-order guarantees differentiability of the solutions.

Having obtained a solution in the domain ${\cal D}$, identified by its value of $V_0$, it can be extended to the $x>0$ interval by using a fourth order Runge-Kutta method with initial value determined by $V_0$.

For the results presented in the next section, we checked that violations of the constraints~\eqref{eqn:algebraic1} and~\eqref{eqn:algebraic2} remain bounded and small over the entire integration domain, which we typically chose to be $x\in[-10,10]$.

\subsection{Results and discussion}
\label{sec:Results}

\begin{figure*}[t]
    \centering
        \subfigure[$\nu=0.450$]{
        \includegraphics[width=0.48\linewidth]{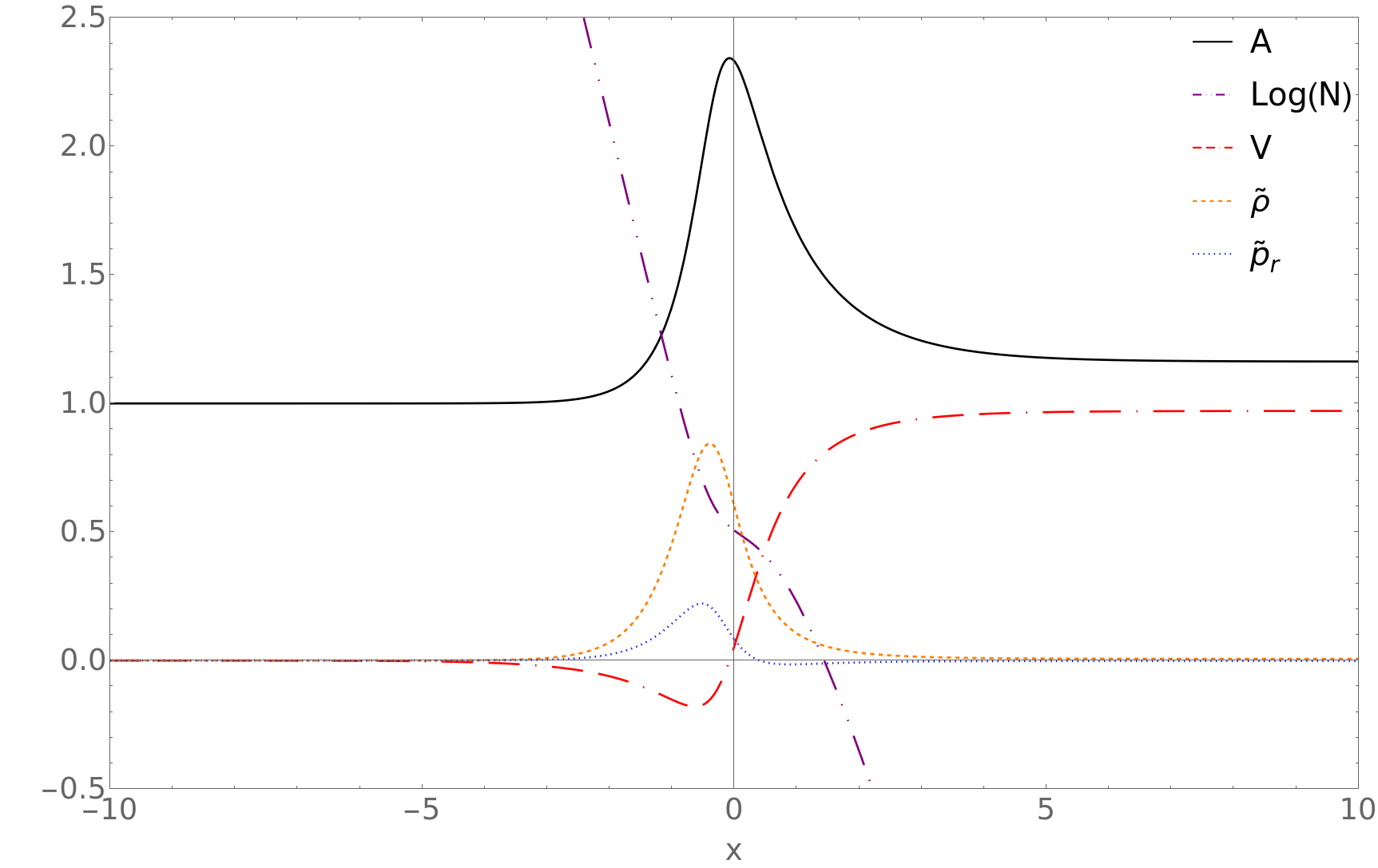}
        \label{fig:quant_sol_450}
    }
    \hfill
    \subfigure[$\nu=0.415$]{
        \includegraphics[width=0.48\linewidth]{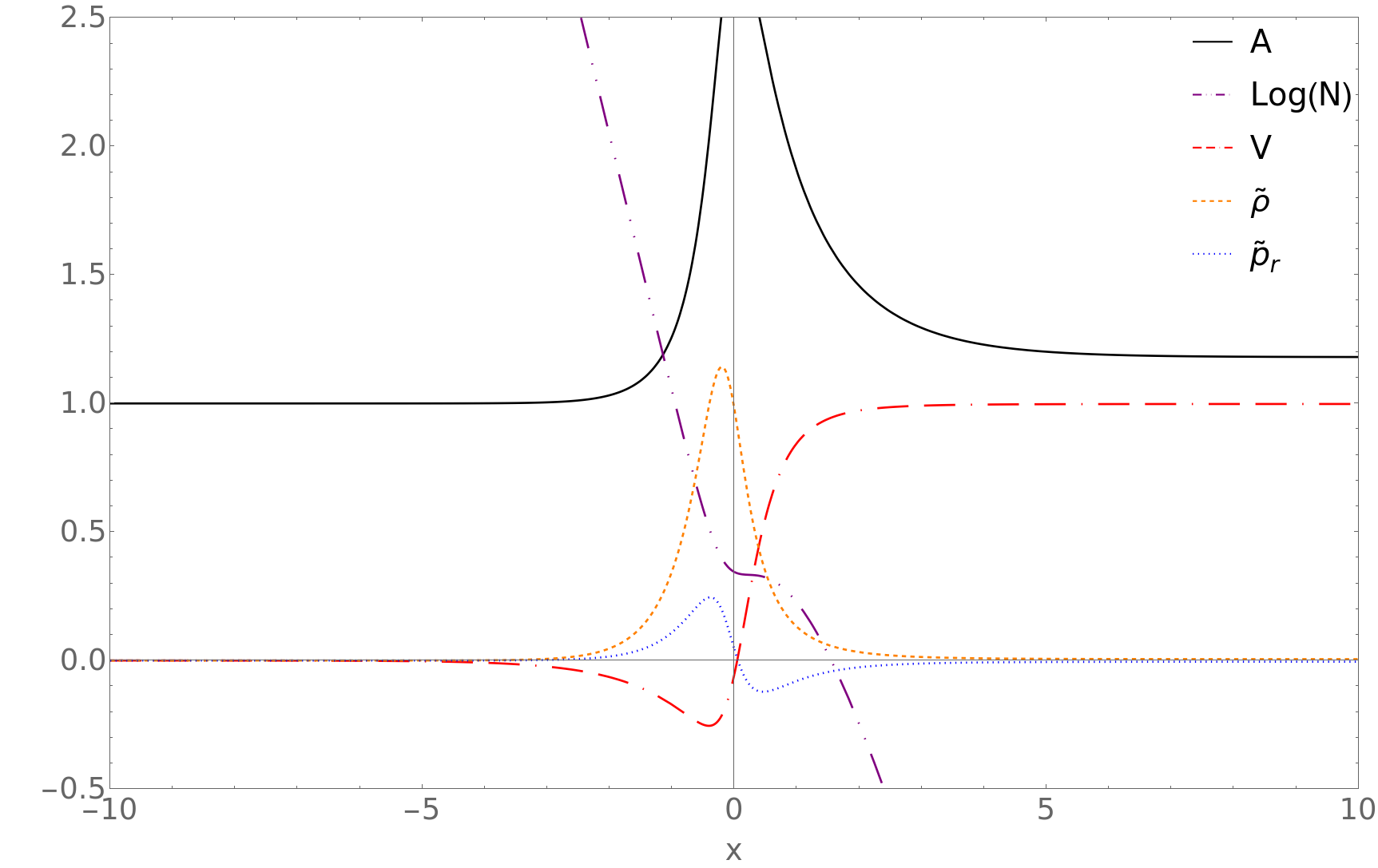}
        \label{fig:quant_sol_415}
    }
    \caption{Spherically symmetric self-similar spacetime with a single sonic point located at $x=0$, for $\n=\s=3$ and two different values of the Poisson ratio: $\nu=0.450$ in panel (a), and $\nu=0.415$ in panel (b). The curves represented are $A$ in solid, $\log{N}$ in dot-dash-dot, $\trho$ in dashed, $\tpr$ in dotted and $V$ in dash-dot-dash.}
    \label{fig:quant_sol}
\end{figure*}


\begin{figure*}[t!]
    \centering
            \subfigure[Profile of the normalized density, $\trho$.]{
        \includegraphics[width=0.48\linewidth]{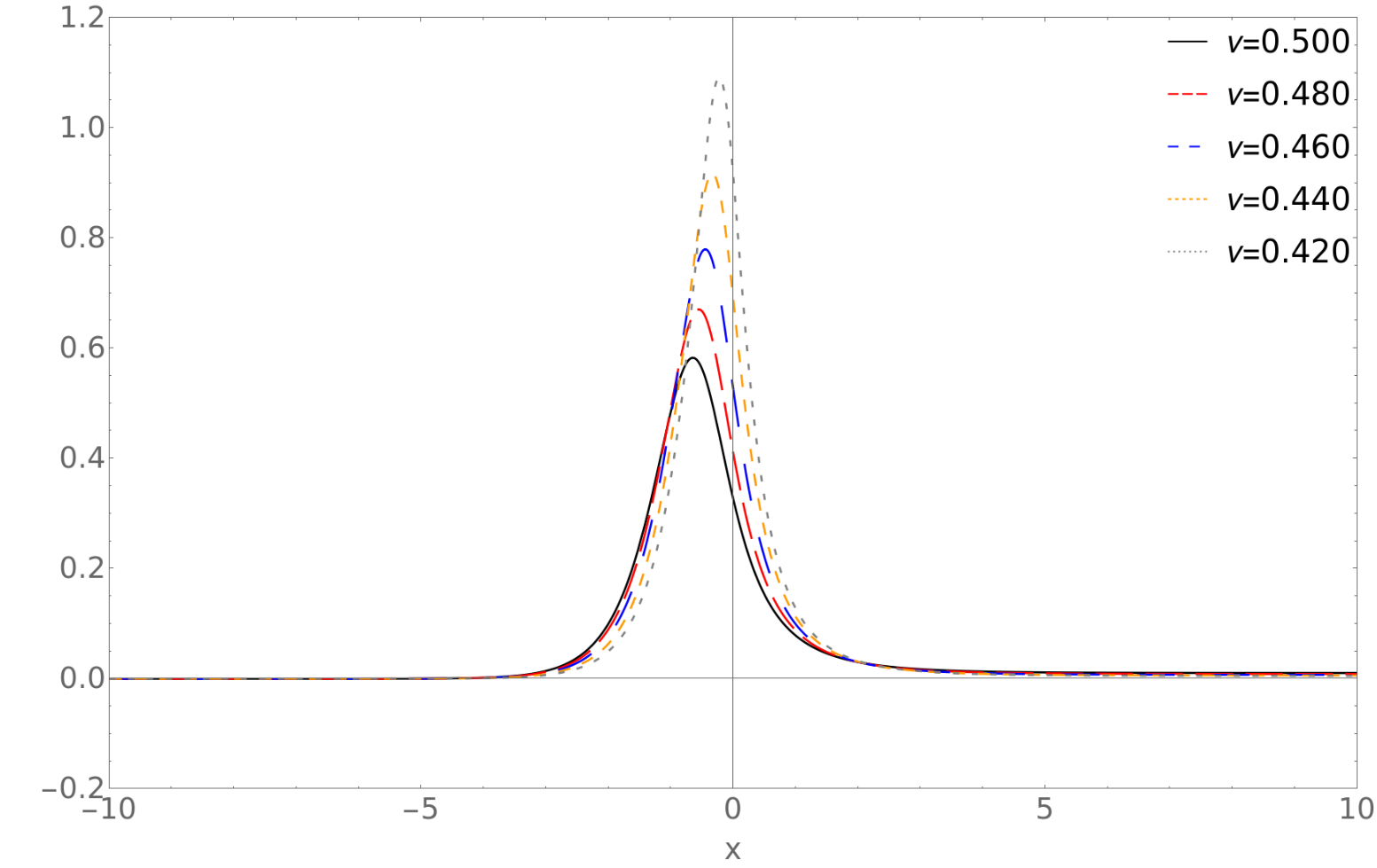}
        \label{fig:rhoset}
    }
    \hfill
    \subfigure[Profile of the normalized radial pressure, $\tpr$.]{
        \includegraphics[width=0.48\linewidth]{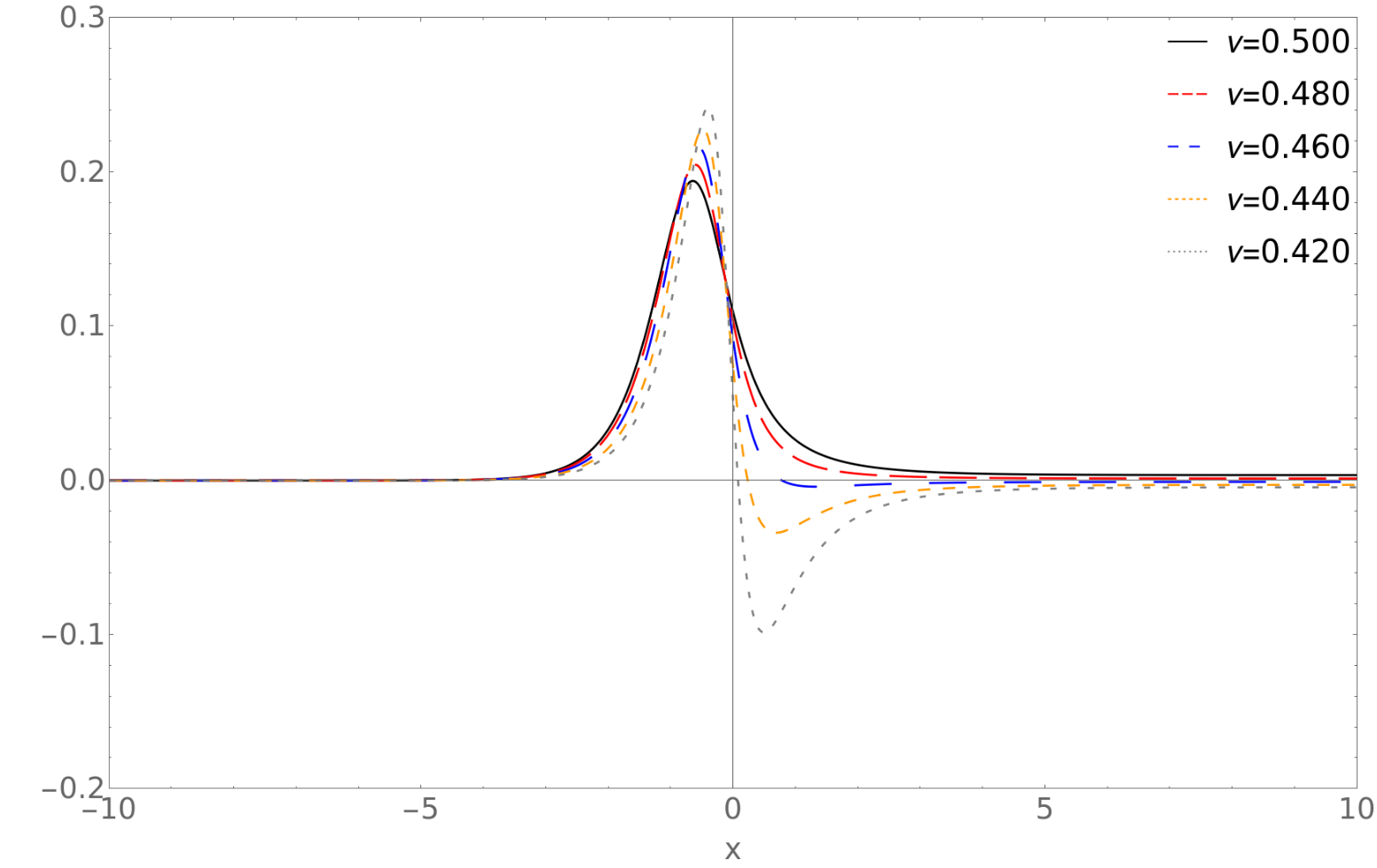}
        \label{fig:prset}
    }
    \caption{Profile of the normalized density, in panel (a), and of the normalized radial pressure, in panel (b), for different values of the Poisson ratio $\nu$, with $\n=\s=3$ fixed. The curves displayed correspond to $\nu=0.500$ (perfect fluid) in solid, $\nu=0.480$ in long-dashed, $\nu=0.460$ in medium-dashed, $\nu=0.440$ in short-dashed and $\nu=0.420$ in dotted.}
    \label{fig:rhoset_prset}
\end{figure*}

Applying the numerical method described above to the radiation fluid elastic counterpart, i.e., $\n=\s=3$, we obtain the results shown in Fig.~\ref{fig:quant_sol_450} for $\nu=0.450$ and in Fig.~\ref{fig:quant_sol_415} for $\nu=0.415$. This solution is analogous to the Evans-Coleman spacetime previously found in the perfect fluid case, and herein generalized to elastic materials. The extension to other values of the Poisson ratio $\nu$ is shown in Figs.~\ref{fig:rhoset} and~\ref{fig:prset} for the profiles of the normalized density and radial pressure, respectively. The profiles of the other functions characterizing the solution are shown in Appendix~\ref{sec:Appendix}. 

While the solution remains qualitatively similar for all values of $\nu$ explored, a sufficiently low Poisson ratio ($\nu\lsim 0.47$) allows the radial pressure to become negative. This is not possible for perfect fluid models as the pressure is related to the density by a positive multiplicative constant factor.
Generally, decreasing $\nu$ generates greater compression near the sonic point.

\begin{figure}[t]
    \centering
    \begin{minipage}[t]{0.48\linewidth}
        \centering
        \includegraphics[width=\linewidth]{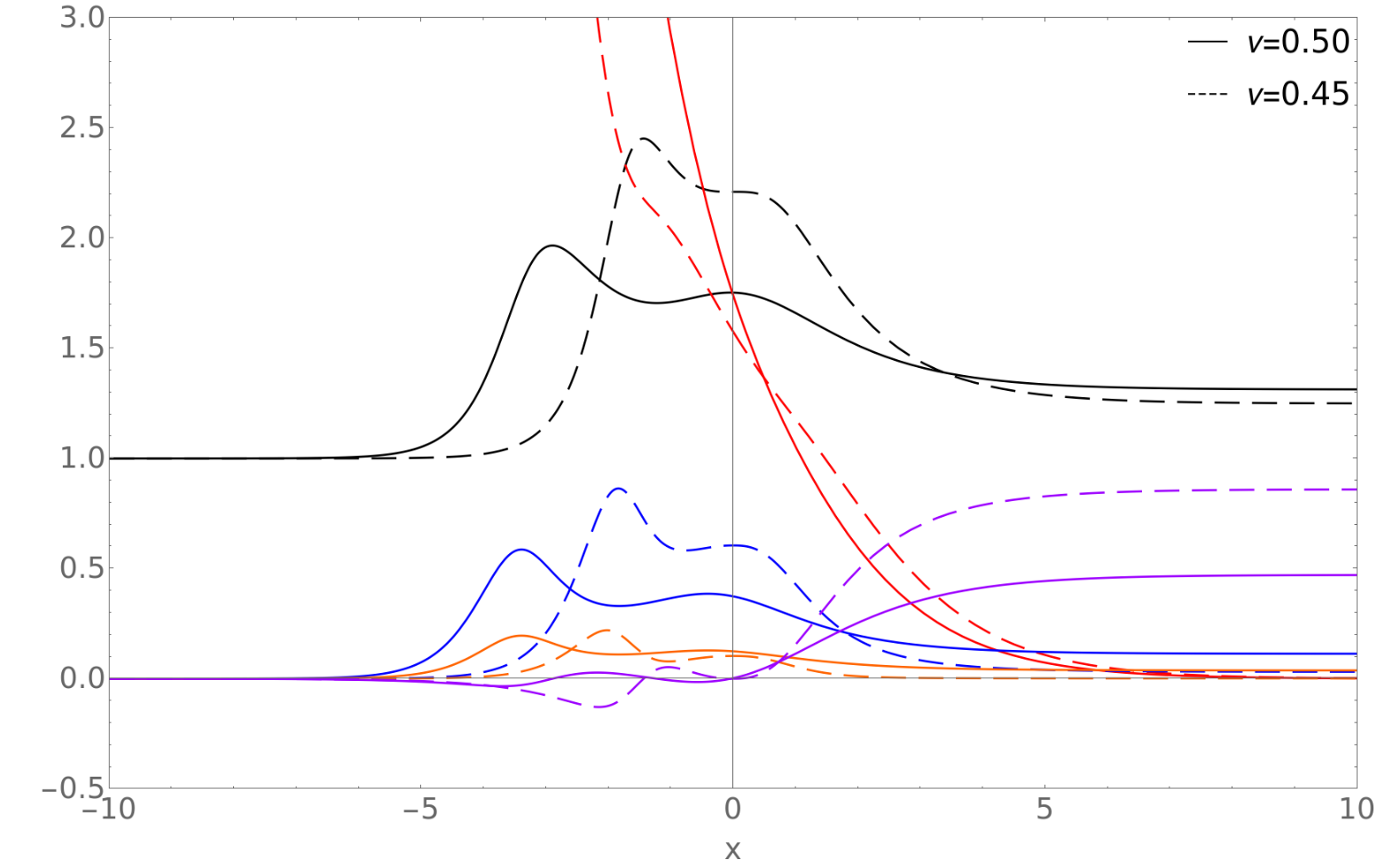}
        \caption{Solution with 3 zeroes for the collapsing spherically symmetric self-similar spacetime with a single sonic point at $x=0$, with $\n=\s=3$ and $\nu=0.50$ in full (perfect fluid limit) and $\nu=0.45$ in dashed. The profiles shown are for $A$, $\log{N}$, $\tilde{\rho}$, $\tilde{p_r}$ and $10\times V$.}
        \label{fig:3zeroes}
    \end{minipage}
    \hfill
    \begin{minipage}[t]{0.48\linewidth}
        \centering
        \includegraphics[width=0.65\linewidth]{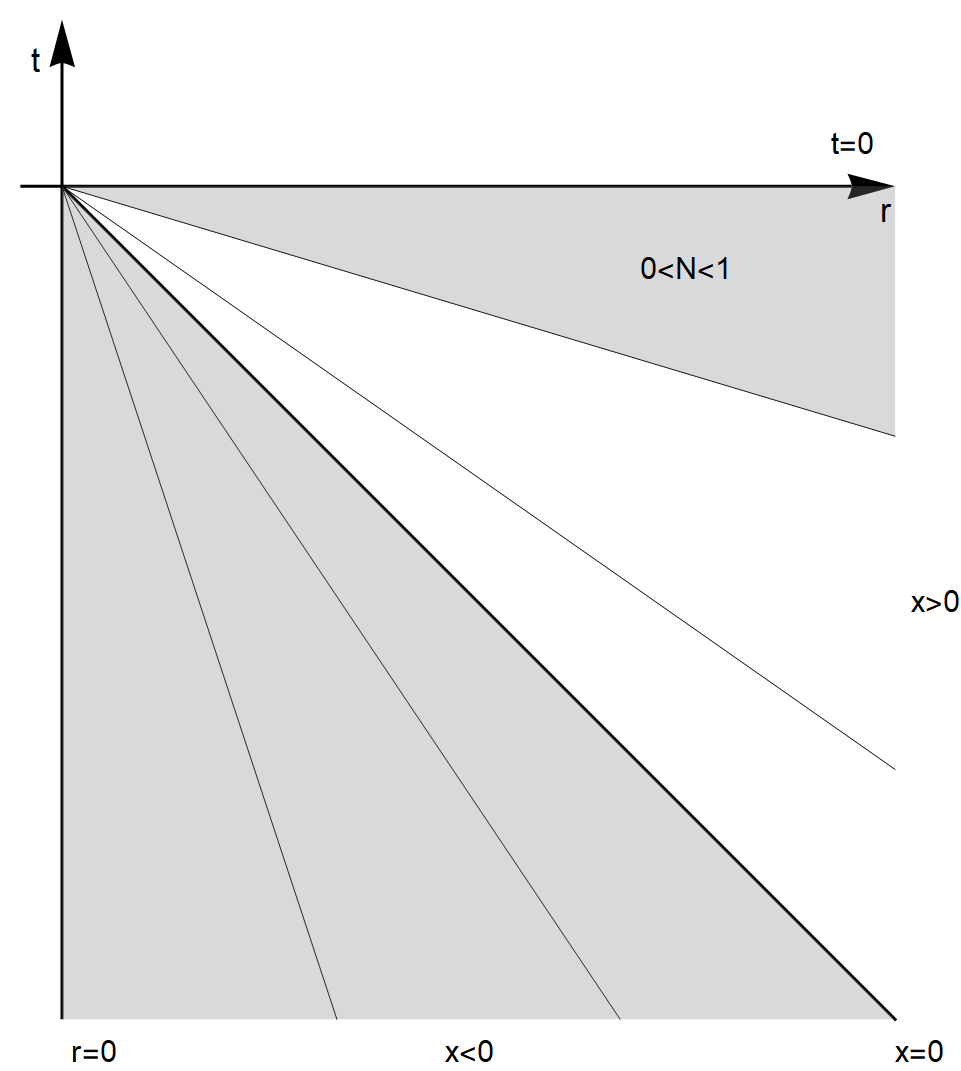}
        \caption{Self-similar spacetime regions. Straight lines represent hypersurfaces of constant $x$ coordinate. For $x<0$, from the regular center to the sonic point, a radial-moving observer at fixed $x$ is timelike and the stellar material falls at a speed lower than that of sound. For $x>0$, beyond the sonic point, there are two regions separated by the $N=1$ condition. The fixed observer is timelike up to it, becoming spacelike to its future.}
        \label{fig:SS_graph}
    \end{minipage}
\end{figure}

The Evans-Coleman solution, and its elastic generalizations, are characterized by the velocity field, $V$, featuring a single zero. This is the fundamental mode. As in the case of the perfect fluid, overtones with multiple zeroes are possible. Fig.~\ref{fig:3zeroes}, shows one such mode with 3 zeroes. Similarly to the fundamental solution, decreasing the Poisson ratio $\nu$ leads to higher gradients in the metric and matter functions.

\medskip
Stellar particles, as seen by an observer at fixed $x$ coordinate, attain superluminal velocities for $x>x_N$, where $x_N>0$ is the smallest value such that $N(x_N)=1$. Recall that for $N<1$ the $(\tau,x)$ coordinates switch character, so at this point the fixed `observer' becomes spacelike, see Fig.~\ref{fig:SS_graph}.
This is a feature common to all the CSS collapses we study, including the perfect fluid limit, but as the Poisson ratio is reduced we observe that $v$ asymptotes smaller (but still superluminal) velocities as $x\to\infty$.

\vfill

\pagebreak

\onecolumngrid 

\twocolumngrid 

\begin{figure*}[t!]
    \centering
    \subfigure[]{
        \includegraphics[width=0.48\linewidth]{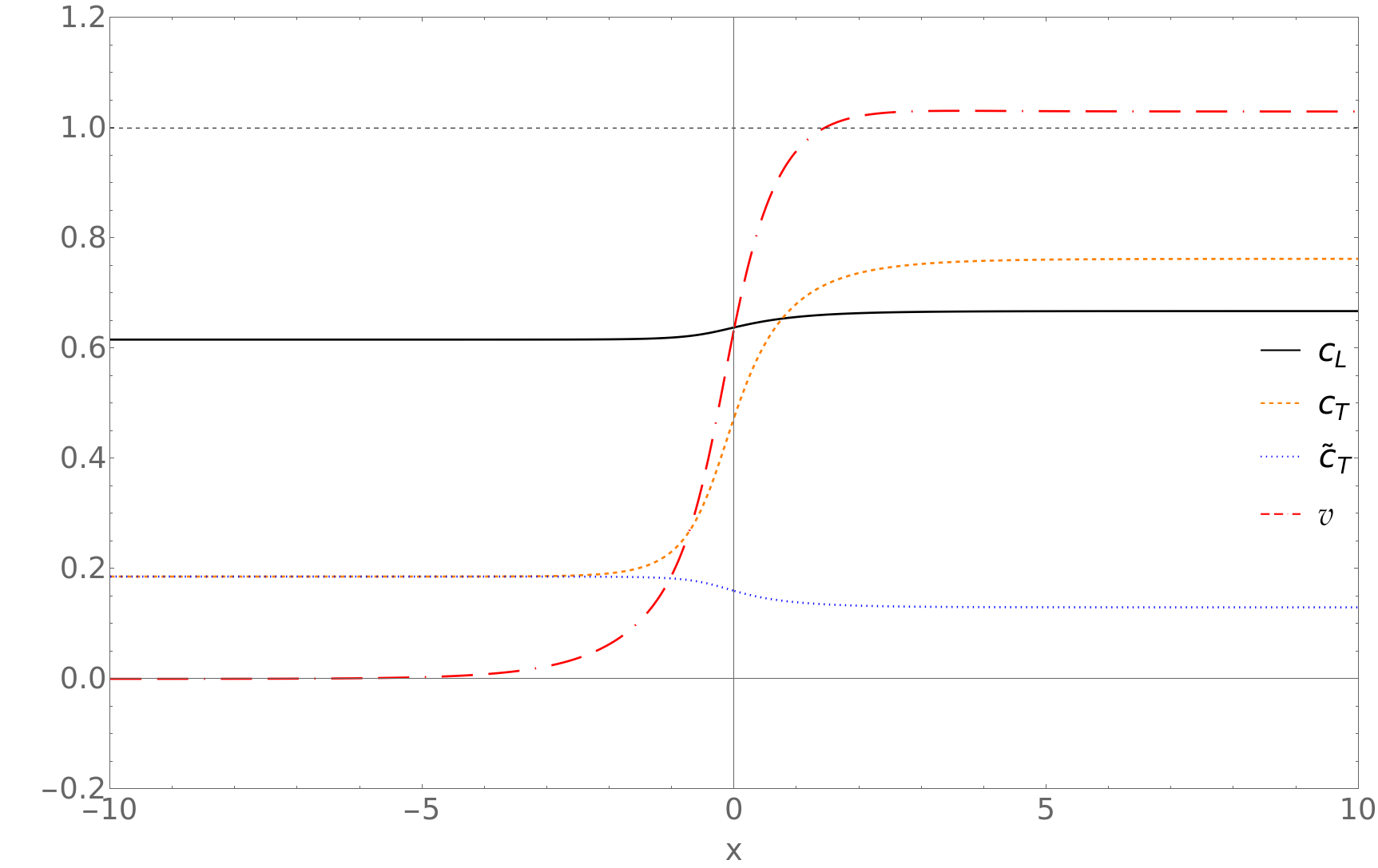}
        \label{fig:vel_sol_450}
    }
    \hfill
    \subfigure[]{
        \includegraphics[width=0.48\textwidth]{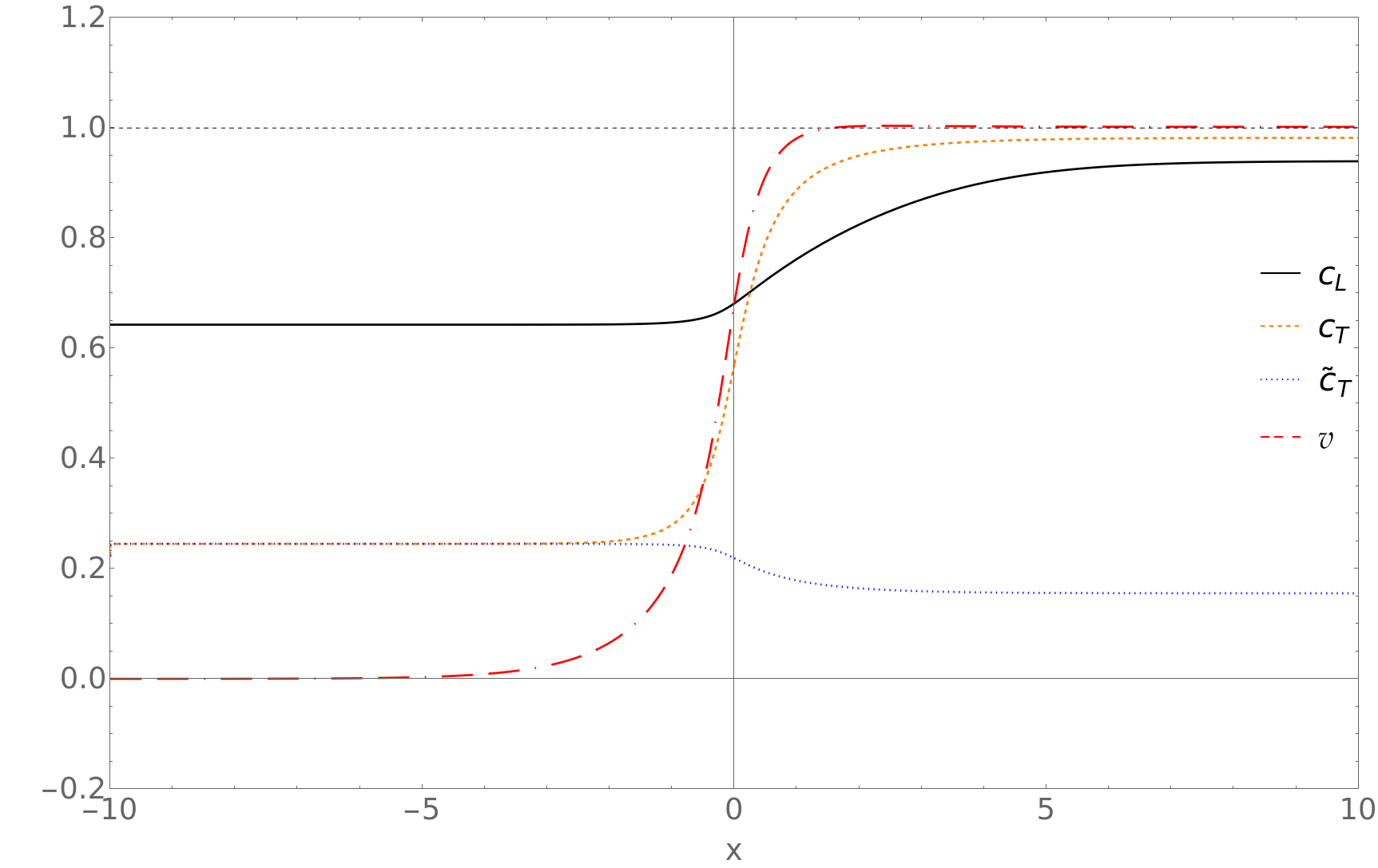}
        \label{fig:vel_sol_415}
    }
    \caption{Spherically symmetric self-similar spacetime with a single sonic point at $x=0$, at $\n=\s=3$ and two different values of the Poisson ratio: $\nu=0.450$ in panel (a), $\nu=0.415$ in panel (b). The curves shown are $c_L$ in solid, $c_T$ in dashed, $\widetilde{c}_T$ in dotted and the velocity observed by the fixed observer, $v$, in dash-dot-dash.}
    \label{fig:vel_sol}
\end{figure*}

~

Another novelty that elasticity introduces is the existence of three independent wave velocities, instead of 
just one, see Fig.~\ref{fig:vel_sol}.
The isotropic wave velocity ---the only one that is supported by perfect fluids--- is identified with the longitudinal radial wave speed, $c_L$, but two additional elastic wave speeds are present: the transverse radial wave speed, $c_T$, and the transverse tangential wave speed, $\widetilde{c}_T$, see Eqs.~(\ref{eq:cT}-\ref{eq:cTtilde}). 
A sonic point occurs when stellar particles propagate at the sound speed $c_L$, as can be seen from Eqs.~(\ref{eq:ab}-\ref{eq:cd}) and \eqref{eq:stellarvelocity}. But unlike the perfect fluid case, this speed is not constant. Indeed, it grows monotonically, albeit at a lower rate than the velocity $v$ in the $N>1$ region. 
This raises the possibility of the occurrence of a second sonic point at some $x>x_N$. Indeed, in our numerical explorations we found this to be the case whenever we lowered $\nu$ below $0.415$, while fixing $\n=\s=3$. As discussed in section~\ref{sec:NumApproach}, these solutions are expected to be singular, and numerically we were unable to obtain regular behavior across the second sonic point, whenever it is present.

\medskip
So far we have kept $\n=\s$ fixed, whilst allowing $\nu$ to vary. However, one can also consider elastic extensions of perfect fluids by detuning the shear index $\s$. In Appendix~\ref{sec:Appendix} we present results obtained when varying $\s$, while keeping $\nu=1/2$ and $\n=3$ fixed. The profiles of the various functions are qualitatively unchanged in this case, with the main effect being an overall rescaling.

\bigskip

\section{Conclusion}

We have shown that relativistic elasticity admits continuous self-similar evolutions. In this context, perfect fluids can be regarded as a limiting case, attainable by fine-tuning the parameters of the elastic model. 
We numerically obtained self-similar solutions for a three-parameter family of a scale-invariant elastic model and analyzed their properties, focusing on departures from the perfect fluid behavior. 
Notably, elasticity allows for negative pressure and higher densities near the sonic point, as well as inducing a non-constant sound speed and the presence of two additional transverse wave speeds.

We have seen that tuning the elasticity parameters sufficiently far from the perfect fluid values leads to the appearance of a second sonic point, and the consequent breakdown of regularity of the solution. Hence, the requirement of analyticity imposes bounds on the elasticity parameters of the material. For example, fixing the polytropic index to $\n=3$ and the shear index to $\s=3$, regular solutions were found only for values of the Poisson ratio satisfying $0.415\lsim\nu\leq1/2$.

The expectation is that the self-similar solutions we have constructed herein ---in particular, the fundamental modes--- correspond to critical solutions in the context of gravitational collapse with elastic materials. 
Naturally, the next step is the study of linear perturbations around these solutions, to confirm if they are indeed the critical solutions and, if so, to extract the critical exponents. This study is under way and will be presented elsewhere.

\pagebreak

Finally, let us remark that the scale-invariant elastic model we adopted, Eq.~\eqref{eq:ScaleInvariantRho}, is just one ---in a sense, the simplest--- such model. Other choices of the free function $h$ also lead to acceptable self-similar elastic evolutions. We leave for future investigation to assess how our present findings change among the viable models.

\acknowledgments
We thank Artur Alho, Tomohiro Harada, David Hilditch, Jos\'e Nat\'ario and Brien Nolan for various useful discussions. Diogo L.F.G. Silva acknowledges support from FCT grant 2022.13617.BD, DOI: https://doi.org/10.54499/2022.13617.BD. JVR and Diogo L.F.G. Silva also acknowledge the support from project 2024.04456.CERN.

\bibliography{refs}

\vfill

\newpage

\begin{appendix}

\onecolumngrid

\newpage
\section{More details about the numerical results}
\label{sec:Appendix}

\begin{figure*}[h]
    \centering
    \subfigure[Profile of the metric function $A$.]{
        \includegraphics[width=0.4\textwidth,height=4cm]{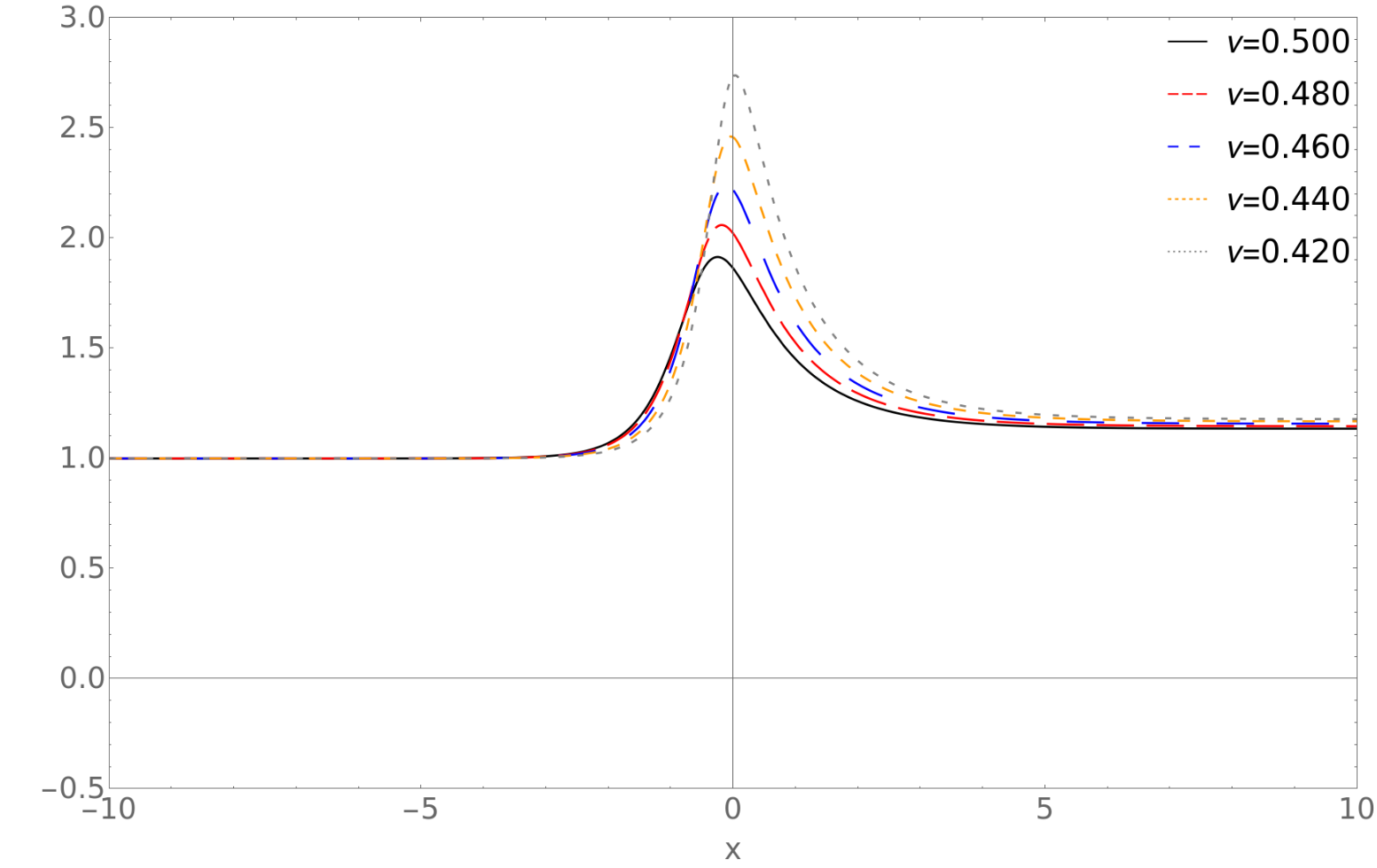}
        \label{fig:Aset}
    }
    \hfill
    \subfigure[Profile of the metric function $\log N$.]{
        \includegraphics[width=0.4\textwidth,height=4cm]{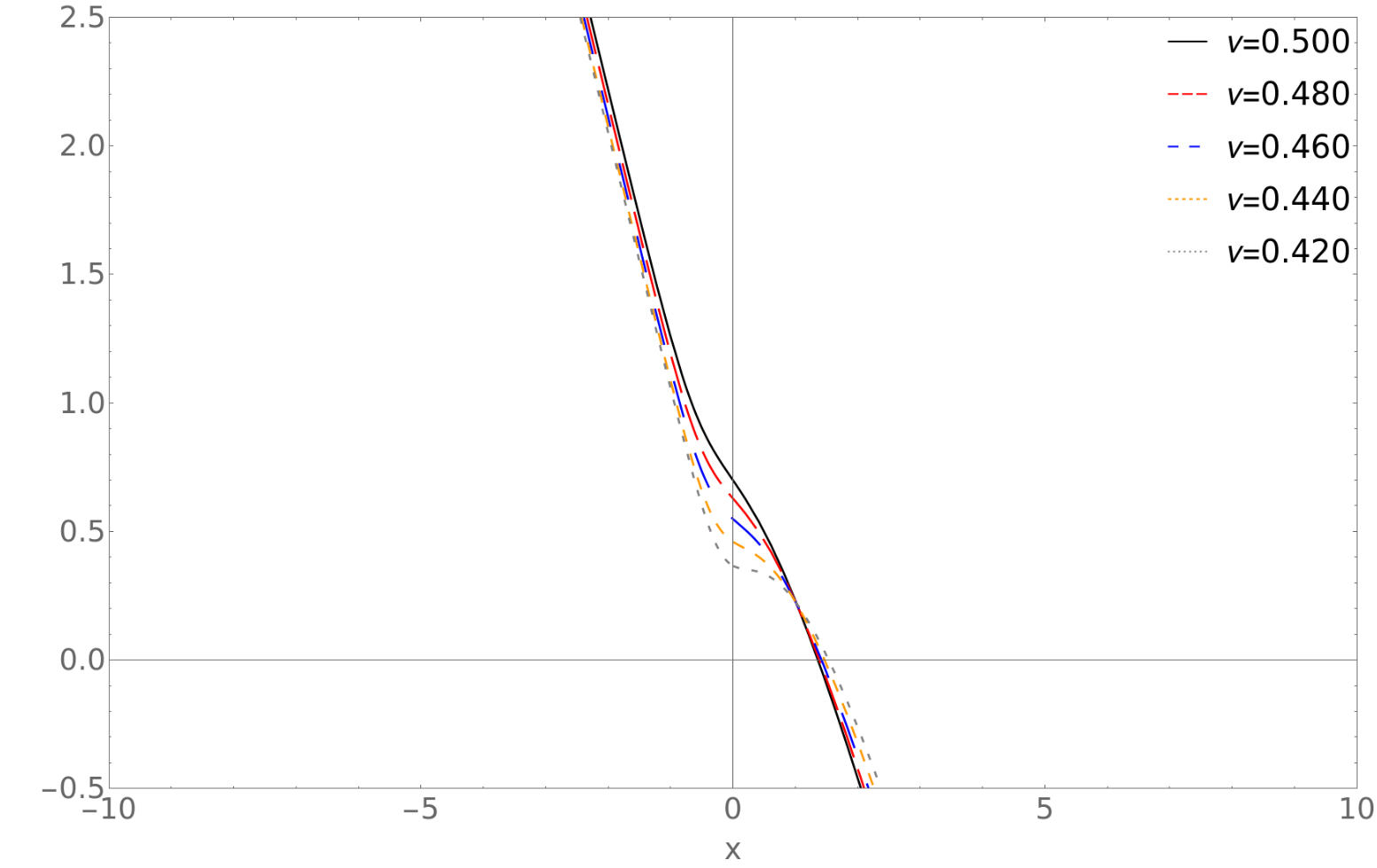}
        \label{fig:Nset}
    }
    \\
    \subfigure[Profile of the 3-velocity $V$.]{
        \includegraphics[width=0.4\textwidth,height=4cm]{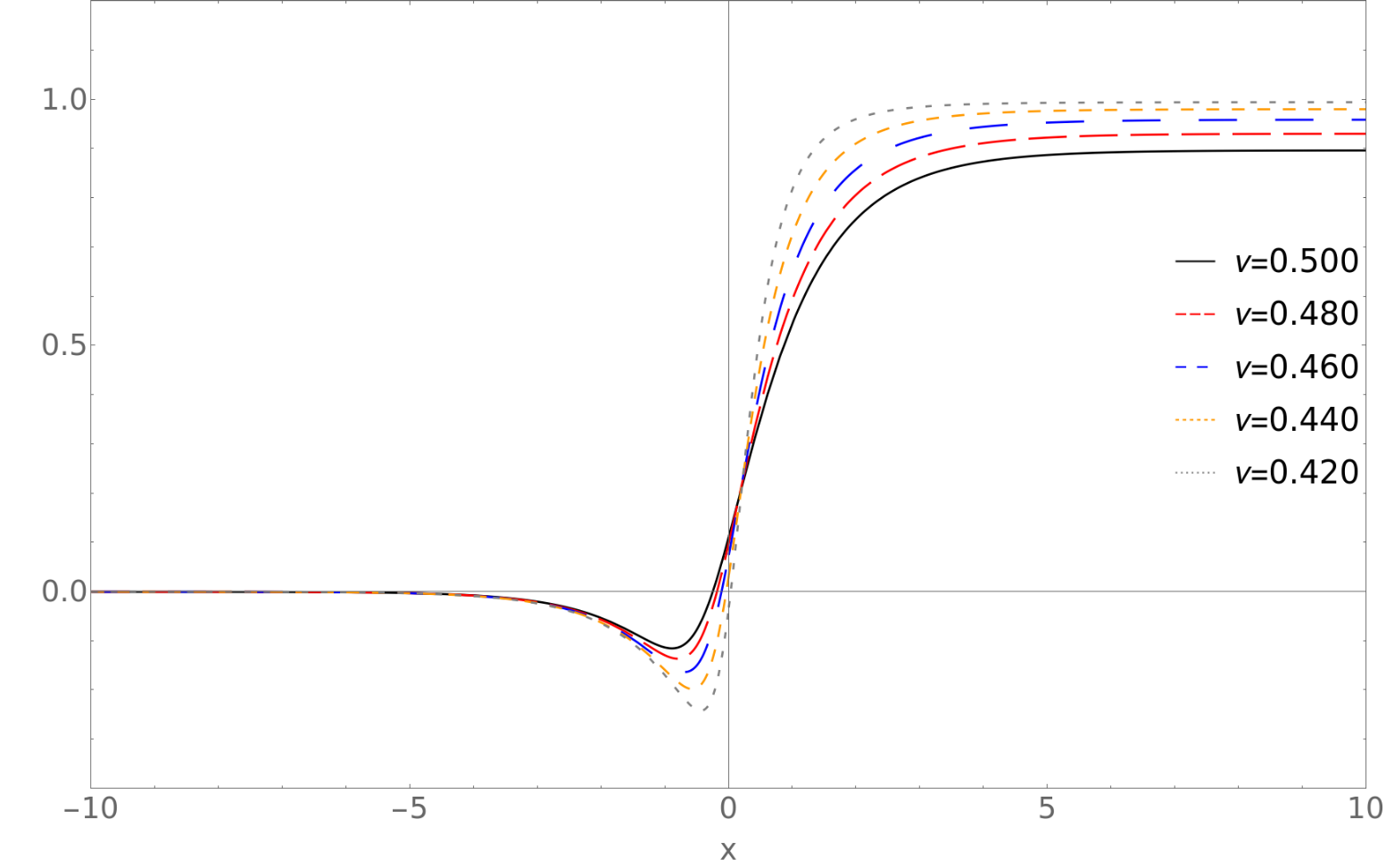}
        \label{fig:Vset}
    }
    \caption{Profiles of functions $A$, $\log{N}$ and $V$ for different values of the Poisson ratio $\nu$, with $\n=\s=3$ fixed. The curves are $\nu=0.500$ (perfect fluid) in solid, $\nu=0.480$ in long-dashed, $\nu=0.460$ in medium-dashed, $\nu=0.440$ in short-dashed and $\nu=0.420$ in dotted.}
    \label{fig:ANVset}
\end{figure*}

The profiles of the normalized density, $\trho$, and the normalized radial pressure, $\tpr$, as a function of the Poisson ratio $\nu$ were shown and discussed in section~\ref{sec:Results}. Here we display the corresponding results for the remaining intervening functions $A$, $N$ and $V$, keeping $\n=\s=3$ fixed, in Fig.~\ref{fig:ANVset}. Once again, the main features in each of these functions are exacerbated as $\nu$ is decreased.

\begin{figure*}[h]
    \centering
        \subfigure[]{
        \includegraphics[width=0.4\textwidth,height=4cm]{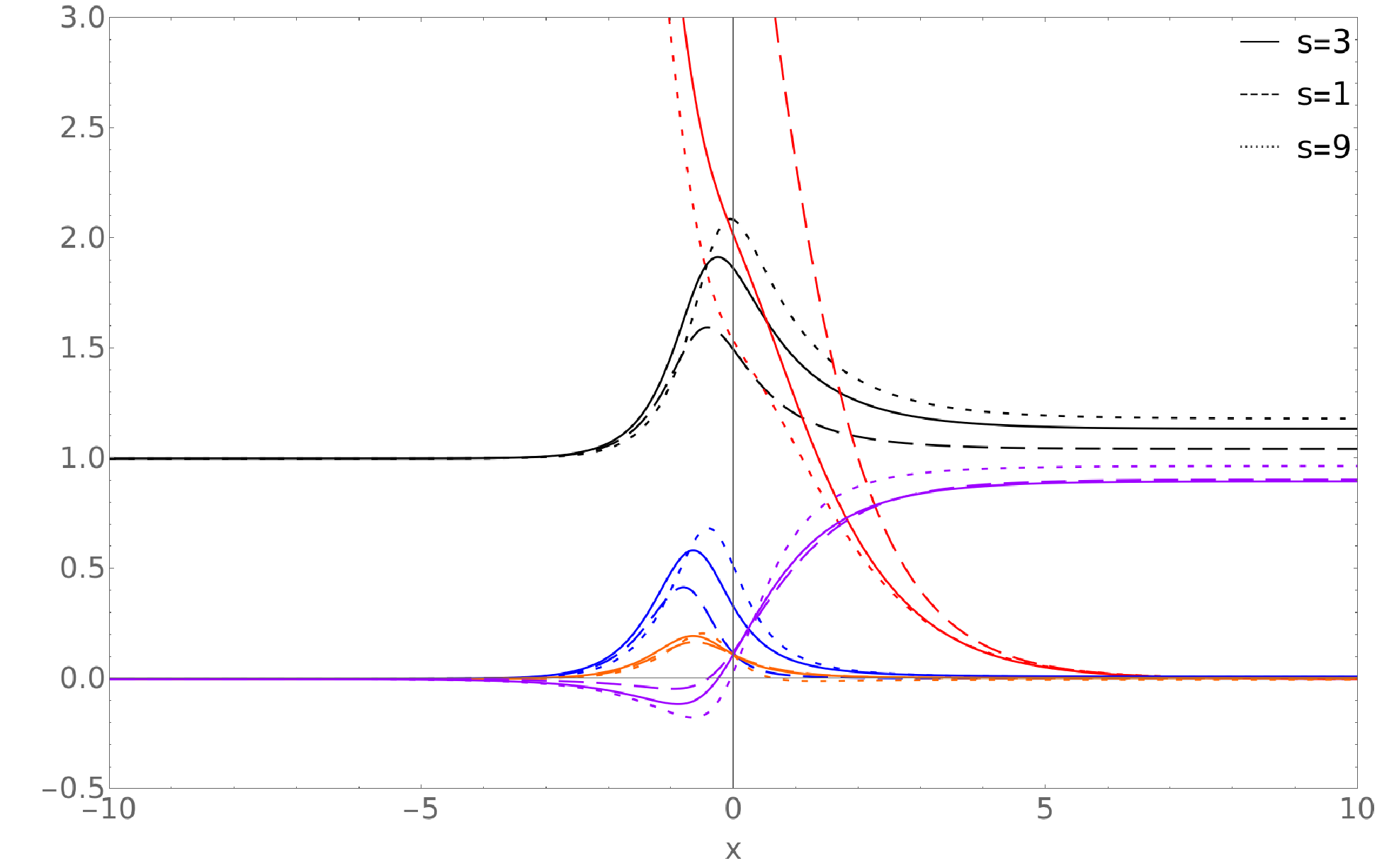}
        \label{fig:varys}
    }
    \hfill
    \subfigure[]{
        \includegraphics[width=0.4\textwidth,height=4cm]{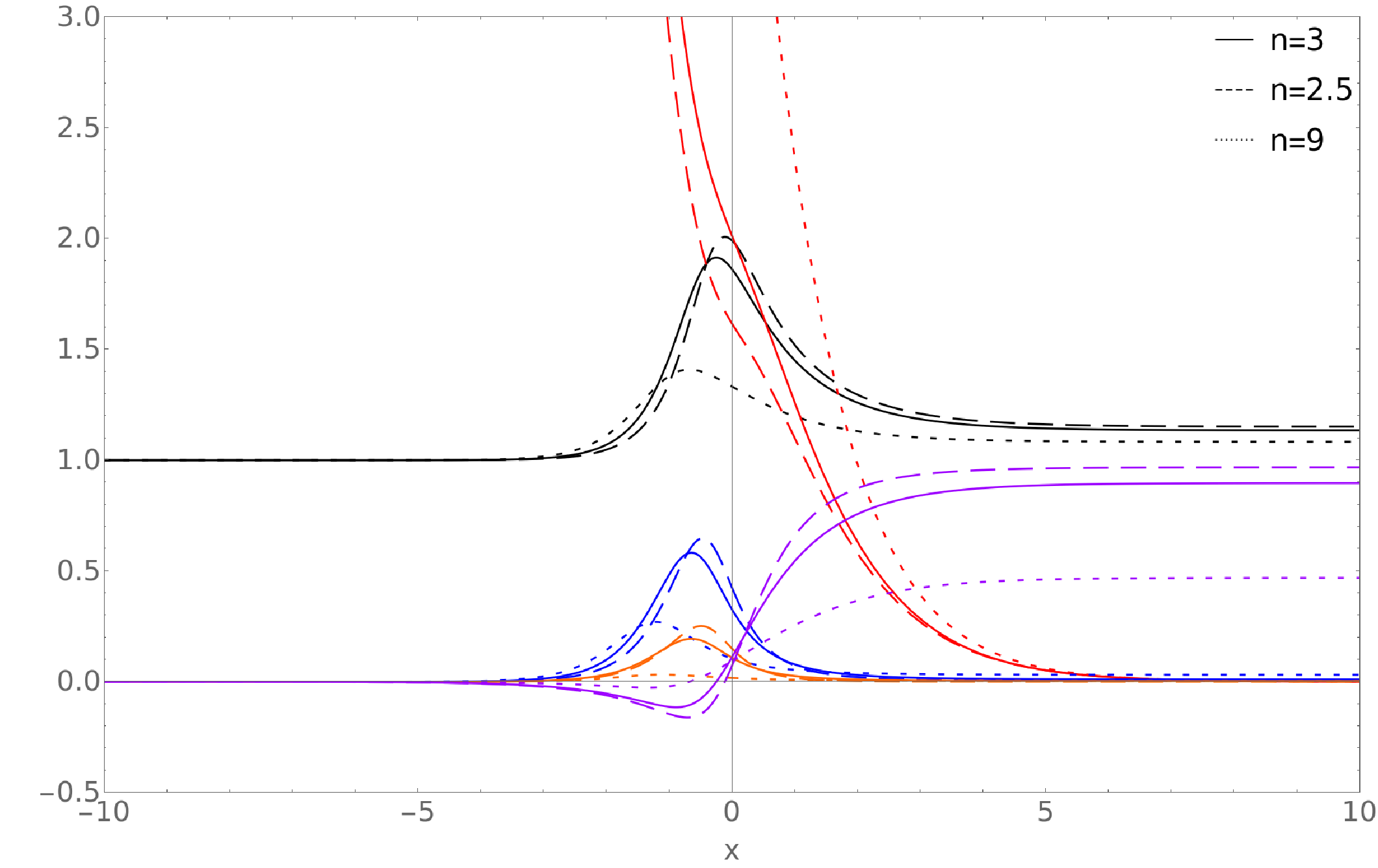}
        \label{fig:Nset}
    }
    \caption{Profile of the $A$, $\log{N}$, $\tilde{\rho}$, $\tilde{p}_r$ and $V$ functions for $\nu=1/2$, varying the polytropic index $\n$ and the shear index $\s$. In panel (a) we fix $\n=3$, showing results for $\s=3$ in full (perfect fluid limit), for $\s=1$ in dashed, and for $\s=9$ in dotted. In panel (b) we fix $\s=3$, showing results for $\n=3$ in full (perfect fluid limit), for $\n=2.5$ in dashed, and for $\n=9$ in dotted.}
    \label{fig:varyn}
\end{figure*}

Besides varying $\nu$, as we mainly discussed in section~\ref{sec:Results}, elastic departures from perfect fluids can also be obtained by changing the shear index, $\s$. 
By keeping $\nu$ fixed and varying $\s$ or $\n$ we find the results shown in Fig.~\ref{fig:varys} and Fig.~\ref{fig:varyn}, respectively. Visual inspection reveals a strong resemblance between the two cases, and shows that changes of either $\s$ or $\n$ mainly produce an overall rescaling of the solution, leaving the profiles qualitatively unchanged. 
It is also found that, should $\s$ become sufficiently larger than $\n$, a new sonic point may appear in the $x>x_N$ region. As discussed in the main text, this would signal the loss of regularity of the solutions.

\end{appendix}

\end{document}